# Sharing Channel In IEEE 802.16 Using The Cooperative Model Of Slotted ALOHA

Abdellah Zaaloul[1], Mohamed Ben El Aattar[1], Mohamed Hanini[1], Abdelkrim Haqiq[1], And Mohammed Boulmalf[2].

[1]Computer, Networks, Mobility and Modeling laboratory FST, Hassan 1st University, Settat, Morocco
e-NGN research group, Africa and Middle East
[2]International University of Rabat

**Abstract**—One of the main problems in WIMAX is to share the medium by multiple users who compete for access. Various random access mechanisms, such as ALOHA and its corresponding variations have been widely studied as efficient methods to coordinate the medium access among competing users. In this paper the slotted ALOHA protocol is implemented as a two-state system. Using Markov Models, we evaluate the channel utilization and we analyze the throughput under different fairness conditions. The cooperative team problem is considered.

## I. INTRODUCTION

In communication network, managing the limited resources attracted big attention in research in the past decades. Indeed, the limited resources in wireless networks lead to several problems that affect the system performance such as delay, energy, throughput and acceptance probability. To overcome this constraint network based on multi-access medium requires mechanisms for effective access to the media.

The IEEE 802.16 (WiMAX) network [1,2,3] provides numerous advantages such as improved performance and robustness end-to-end IP-based networks, secure mobility and broadband speeds for voice, data, and video. It leads a very high utilization of radio resources and a good quality of service (QoS) framework. However, In 802.16 (WiMAX), at network entry, subscriber stations (SS) contend for resource on the initial ranging interval in order to access the network [4]. This contention based access scheme at the medium access control (MAC) layer in 802.16 WMANs is expected to be the main mode of operation for supporting the QoS requirements, especially for the best effort (BE) class of traffic generated by most Internet applications (web surfing, FTP, etc.).

In wireless networks various techniques can be used to reduce collisions in a contention situation. For example, the Ethernet uses CSMA-CD (Carrier Sense Multiple Access with Collision Detection) as a MAC protocol, while 802.11 wireless LAN uses CSMA-CA (Carrier Sense Multiple Access with Collision Avoidance). But in wireless communication settings, collision detection (CD) is expensive and collision avoidance through carrier sensing is often difficult for high-density, low-cost devices. An alternative, lightweight protocol that has been often used to ovoid such a problem is the Slotted ALOHA (S-ALOHA), used particularly in IEEE 802.16 [5]

The slotted-ALOHA protocol [6] was introduced to improve the utilization of the shared medium by synchronizing the transmission of devices within time-slots because the ALOHA protocol [7] is a fully decentralized medium access control protocol that does not perform carrier sensing. Recently, various forms of the slotted-ALOHA protocols are used in most of the current digital cellular networks such WiMAX and GSM (Global system for Mobil Communication) where the control channels of the TDM channels use this mechanism.

For our analysis, we consider the situation where users (wireless nodes) communicate over a random access channel. That is, whenever a user (node) has a new packet to send, it will do so immediately. If packets are sent simultaneously by more than one user then they collide. After the end of the transmission of a packet, the transmitter receives the information on whether there has been a collision (and retransmission is needed) or whether it was well received. All packets involved in a collision are assumed to be corrupted and are retransmitted after some random time (a backoff algorithm).

In this work we assume that the entire slot is used solely by the BE (Best Effort) traffic in a WiMAX network. Taking this assumption into consideration, BE uses choice of retransmission probabilities to obtain access to the service. Multiple transmissions simultaneously result in a collision. We model a system of m users implementing a cooperative slotted ALOHA protocol with tunable parameters via Markov Models that allow us to measure nodes rates success (throughput) under different fairness conditions.

Cooperation is in the sense that all users retransmit with the same probability, and fairness is ensured by the fact that the time of occupation of the channel by a user once it transmits a packet with success has the same distribution for all users. such that users will achieve the same performance on average.

The rest of the paper is organized as follows: We begin by introducing a brief overview of related of related work in section 2, in section 3 we construct a Markov Model and we measure the system throughput in a cooperative team problem where users want to maximize the total throughput of the system. In section 4 we discuss numerically the throughput and delay under different fairness conditions. Section 5 will be devoted to the conclusion.

## II. RELATED WORK

A number of multiple access protocols have been proposed in the literature. The most popular protocols are those based on slotted-ALOHA [8]. These protocols have been adopted mainly by broadband access networks such as in satellite as well as cellular telephone networks for the sporadic radio channels [9], and most recently in the IEEE 802.16 standard. In this protocol, a user with a data session follows a slotted ALOHA contention mechanism to access the media. A user transmits the first packet of the data session by contention to access the media. Once it successfully accesses a slot, it reserves the same slot in future frames until the end of data session where that slot is released [10]. In [11,12] the slotted ALOHA exhibits an instability, namely, in this protocol, the number of backlogged users with packets awaiting to be retransmitted is steadily growing; research in [13] focused on stabilization.

Rivest proposed in [12] a pseudo-Bayesian algorithm, he utilizes feedback to estimate the number of current backlogged nodes in the system. In [14] a Markovian decision model is formulated for dynamic control of unstable slotted ALOHA protocol and optimum decision rules are found. Authors in [15] analyze the stability properties of slotted ALOHA with capture for random access over fading channels with infinitely-many users; Their analysis shows that, with regular users, the system is unstable under any kind of power and probability control mechanism that is based only on decentralized channel state information. Another approach adopted is the implementation of an admission control process that limits the number of simultaneous users in the system to avoid stabilization challenges.

In the last years, the slotted ALOHA protocol is still a current topic in scientific research. Thus, Patet et al [16] have analyzed the number of backlogged packets by using statistical approach and stabilize the expected number of backlogged packets minus mean number of packets which are successfully transmitted.

The authors in [17] proposed an adaptive Slotted ALOHA Algorithm that can accelerate the adjusting speed and can acquire stable throughput on the conditions that there is large fluctuation of system load.

The authors in [18] propose a study based on the Markov chain model, to define optimal binomial distribution probabilities of retransmission and arrival packets, in the case of slotted ALOHA protocol; they defined the average packet delay and the throughput, and show the effect of the increase of the number of sources on these parameters.

In [19] a repeated Bayesian slotted ALOHA game model to analyze the selfish behavior of impatient users is proposed. The authors prove the existence of Nash equilibrium mathematically and empirically. The proposed model enables any type of transmission probability sequence to achieve Nash equilibrium without degrading its optimal throughput. Those Nash equilibria can be used as a solution concept to thwart the selfish behaviors of nodes and ensure the system stability.

In this paper, we argue why time-based fairness is desirable in some cases and we analyze the achieved throughputs of competing nodes, possibly using different data rates and packet sizes in 802.16 cellular networks. We are interested the performance evaluation, in the case of state of the shared channel by m users. We provide a simple model that represents the time of occupation of the channel by a user once it transmits a packet with success.

## III. MODELING UPLINK CHANNEL UTILIZATION

To control the medium access, MAC protocol coordinates the nodes in a network and resolves the contention among their accessing the shared medium, so that the resources are shared fairly and efficiently [7]. In this section, we construct a Markov Model from which we can analyze the throughput and we describe cooperative slotted-ALOHA MAC protocol in which time is divided into units. At each time unit a packet may be transmitted, next in this protocol each node is in either a backlogged state or a free state. There for the decision is actually Markovian for each node in slotted - ALOHA. We will use a two system state as the following: one state is the busy state when only one of the users is transmitting; the other state is either when the system is idle or when collisions happen. We adopt the following notations for a generalized Markov model for slotted- ALOHA type MAC protocol.

Throughout we use the following notations:

$m$: Number of users in the system.

$P_t^i$: Transmitting probability at free states for user i.

$P_r^i$: Transmitting probability at backlogged states for user i.

$Th_i$: Throughput function, which indicates the average throughput of user i.

For all users i, we can assume that $P_t^i = P_t$ and $P_r^i = P_r$

The transition probabilities are:

$P_0 = mP_r(1 - P_r)^{m-1}$ this indicates the probability that only one of the m users transmits.

$P_c = (1 - P_r)^{m-1}$ Which indicates the probability that all of the m-1 users (backlogged nodes) doesn't transmit. Fig.1 shows the Markov chain for two states. The transition matrix for the above Markov chain is:

$$P = \begin{pmatrix} P_c & 1 - P_c \\ P_0 & 1 - P_0 \end{pmatrix}$$

The steady state distribution is solution of the following problem:

$$\pi = \pi P \quad \text{with} \quad \pi = \begin{pmatrix} \pi_1 \\ \pi_2 \end{pmatrix} \text{ and } \pi_1 + \pi_2 = 1$$

We obtained the system:

$$(S) \begin{cases} \pi_1 * P_c + \pi_2 * P_0 = \pi_1 \\ \pi_1 * (1 - P_c) + \pi_2 * (1 - P_0) = \pi_2 \\ \pi_1 + \pi_2 = 1 \end{cases}$$

We find $\pi_1 = \frac{P_o}{1-P_c+P_o}$ and $\pi_2 = \frac{1-P_c}{1-P_c+P_o}$

**Result (a):**

We prove that the throughput is close to $\frac{m}{2m-1}$ when $P_t = 1$ and $P_r \to 0$.

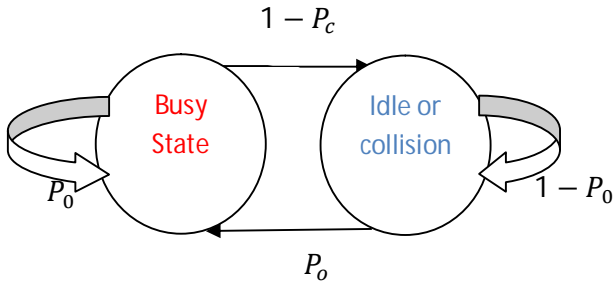

**Figure 1: m users Markov chain with $P_t = 1$ and $P_r \to 0$**

**Proof**:

Let $\pi_1$ be the throughput for the system: $\pi_1 = \frac{P_o}{(1-P_c+P_o)}$

$$= \frac{mP_r(1-P_r)^{m-1}}{1-(1-P_r)^{m-1}+mP_r(1-P_r)^{m-1}}$$

$$= m / \left(\frac{1-(1-P_r)^{m-1}}{P_r(1-P_r)^{m-1}} + m\right)$$

$$= m / \left(\frac{1+(1-P_r)+(1-P_r)^2+(1-P_r)^3+\cdots+(1-P_r)^{m-2}}{(1-P_r)^{m-1}} + m\right)$$

Therefore, $\pi_1 \to \frac{m}{2m-1}$ as $P_r \to 0$.

## IV. EVALUATION PERFORMANCE

*A. Retransmission probability*

In general, after one user successfully transmits a packet, it obtains the channel. This user will continue to occupy the channel for a random amount of time T. We assume that T is an exponential random variable with parameter $\lambda = (1-P_c)$. Let $E[T]$ the mean for the random variable T. If we put $E[T]=U$.

We know that $E[t] = \frac{1}{\lambda}$

$U = \frac{1}{1-P_c}$

$U = \frac{1}{1-(1-P_r)^{m-1}}$

*B. Throughput:*

Using the above results, the total throughput can be computed as a function of U, it is given by:

$$\pi_1 = \frac{mP_r(1-P_r)^{m-1}}{[1+(mP_r-1)(1-P_r)^{m-1}]}$$

$$= \frac{m(U-1)}{\left[m(U-1) + 1/\left(1-\left(1-\frac{1}{U}\right)^{m-1}\right)\right]}$$

Fig.2 shows that the throughput decreases when the retransmission probability increases, that the system becomes backlogged, as there will be more probability that collision occurs. When this probability becomes close to 1 the system throughput is almost zero. And for different values of m when $P_r \to 0$ the throughput values are close to $\frac{m}{2m-1}$ which is proved above.

For m=10 for example, to keep the level of throughput up to 15% and to assure a fairly sharing of the available bandwidth the retransmission probability should not exceed 0.3.

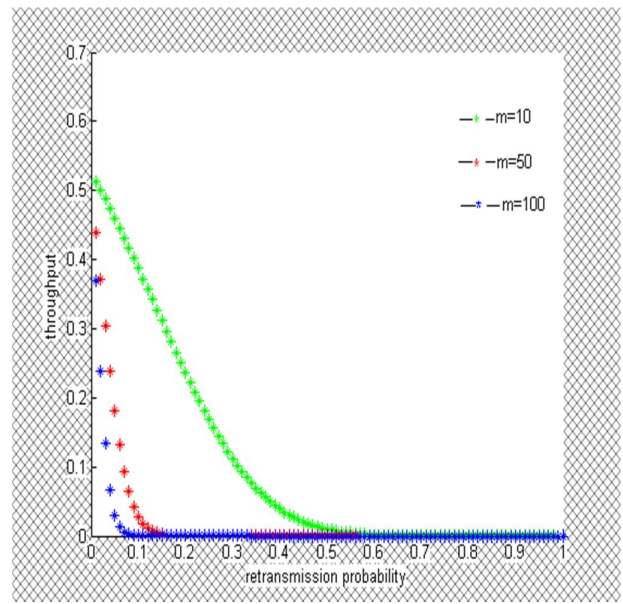

**Figure 2: throughput vs Pr.**

In Fig.3, we can observe that we can achieve very good short-term fairness, without sacrificing much throughput. If we want for example the system to be 24-short-term fair, we Can achieve a total throughput close to 0.5 even for large values of m. Actually, the total throughput does not collapse to zero, even if $m \to \infty$, see the result (a).

When U decreases (the time occupancy of the system by a user who had a success), we note that the system throughput decreases. Indeed, in this case backlogged users retransmit and probability of collision increases.

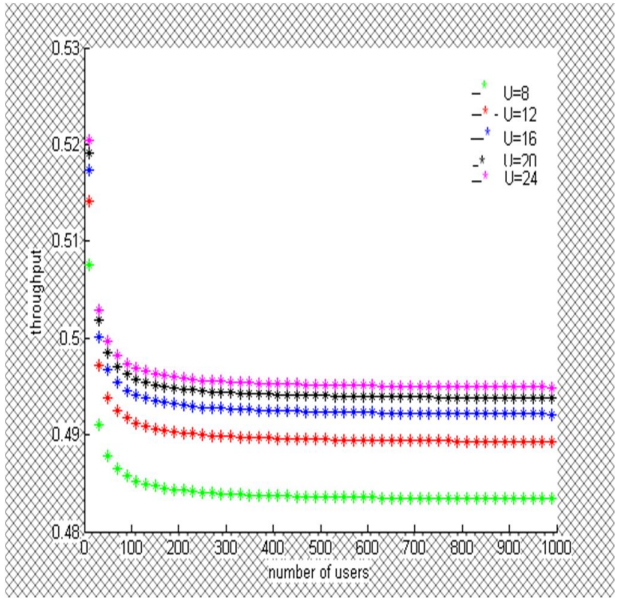

**Figure 3: throughput vs users and under different values of U.**

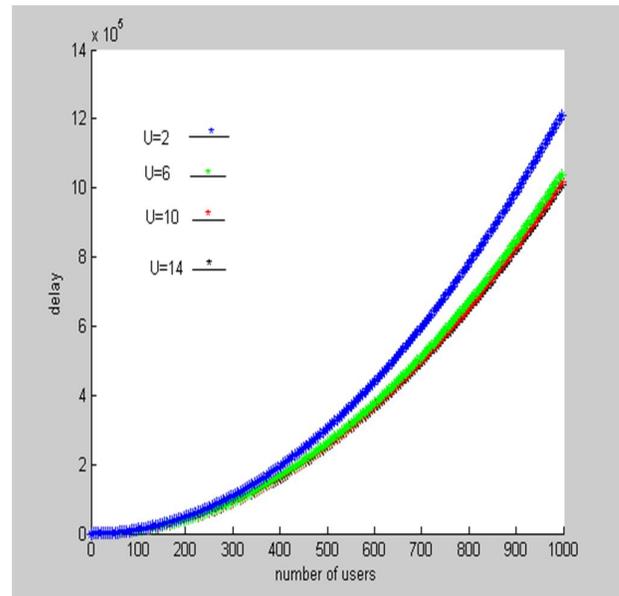

**Figure 4: Delay under different fairness conditions.**

*C. Delay:*

Let us now to analysis the Delay for the system. We note by Q the number of users in the system.

$$Q = \pi_1 * n_1 + \pi_2 * n_2$$

Where $n_1 = \sum_{i=1}^{m} i$ represents the number of users in the cases where the system can be in busy state.

And $n_2 = \sum_{i=2}^{m} i$ represents the number of users in the cases where a collision is probable (the system can be in a idle or collision state).

$$Q = \pi_1 * \frac{m * (m + 1)}{2} + \pi_2 * \frac{(m - 1) * (m + 2)}{2}$$

Therefore by the formula of LITTLE [20] the delay is:

$$D = \frac{Q}{\pi_1}$$

We find $D = \frac{m(m+1)}{2} + \frac{(m-1)(m+2)}{2m\left(1-\sqrt[m-1]{1-\frac{1}{u}}\right)(u-1)}$

Figure 4 shows the total delay under different values of U and m. under low load whatever the value U, the delay is low. Therefore if the load increases the delay also increases, but for higher values of U the delay is lower.

## V. CONCLUSION

In this work, we studied the cooperative slotted-ALOHA protocol at two- states. We constructed a Markov model. And we got the results that are possible to achieve a desired system performance by suitably choosing the mean time channel occupancy **U**. In addition, we assumed that the users cooperate to fairly share the available bandwidth and to maximize the aggregate throughput. An aggregate throughput of at least one half can be achieved even if the number of competing users for the bandwidth increases. The analysis of another performance which is the delay shows that the delay can be minimized for well chosen values of U.

One important feature of this work is to consider the extension of the analysis to delay-sensitive traffic (UGS, rtPS classes), but this will pose additional challenges that need to be carefully evaluated.